\documentstyle[12pt]{article}
\begin{document}
\baselineskip=0.7cm
\newcommand{\EQ}{\begin{equation}}
\newcommand{\EN}{\end{equation}}
\newcommand{\EQA}{\begin{eqnarray}}
\newcommand{\EQN}{\end{eqnarray}}
\newcommand{\e}{{\rm e}}
\newcommand{\Sp}{{\rm Sp}}
\renewcommand{\theequation}{\arabic{section}.\arabic{equation}}
\newcommand{\Tr}{{\rm Tr}}
\renewcommand{\thesection}{\arabic{section}.}
\renewcommand{\thesubsection}{\arabic{section}.\arabic{subsection}}
\makeatletter
\def\section{\@startsection{section}{1}{\z@}{-3.5ex plus -1ex minus
 -.2ex}{2.3ex plus .2ex}{\large}}
\def\subsection{\@startsection{subsection}{2}{\z@}{-3.25ex plus -1ex minus
 -.2ex}{1.5ex plus .2ex}{\normalsize\it}}
\def\appendix{
\par
\setcounter{section}{0}
\setcounter{subsection}{0}
\def\thesection{\Alph{section}}}
\makeatother
\def\thefootnote{\fnsymbol{footnote}}
\begin{flushright}
EFI-98-24 \\
UT-KOMABA/98-15\\
June 1998
\end{flushright}
\vspace{1cm}
\begin{center}
\Large
Short-Distance  Space-Time Structure \\
and Black Holes in String Theory : \\
A Short Review of the Present Status
\footnote{
To appear in the special issue of
the {\it Journal of Chaos, Solitons and Fractals}
on "Superstrings, M,F,S...Theory".
}

\vspace{1cm}
\normalsize
{\sc Miao Li}
\footnote{
e-mail address:\ \  {\tt mli@curie.uchicago.edu}}

\vspace{0.3cm}
{\it Enrico Fermi Institute, University of Chicago\\
5640 Ellis Avenue, Chicago, IL 60637

\vspace{0.5cm}

{\sc Tamiaki Yoneya}
\footnote{
e-mail address:\ \ {\tt tam@hep1.c.u-tokyo.ac.jp}}
\\
\vspace{0.3cm}

 {\it Institute of Physics, University of Tokyo\\
Komaba, Meguro-ku, 153 Tokyo}}

\vspace{1cm}
Abstract
\end{center}

We briefly review the present status of string theory
from the viewpoint of its implications on
the short-distance space-time structure  and
black hole physics.  Special emphases are given on two
closely related issues
in recent developments towards
nonperturbative string theory, namely, the role of
the space-time uncertainty relation
as a qualitative but universal characterization of the
short-distance structure of string theory and
the microscopic formulation of black-hole entropies.
We will also suggest that
 the space-time uncertainty relation
can be an underlying principle for the holographic property
of M theory, by showing that the space-time
uncertainty relation naturally explains the UV/IR relation 
used in a recent derivation of the holographic bound
for D3 brane by Susskind and Witten. 

\newpage
\section{Historical introduction}

String theory emerged originally from the discovery of
the Veneziano formula \cite{veneziano} for the scattering
amplitude for mesons, satisfying the
so-called {\it s-t} duality in an idealized form.
The {\it s-t} duality says that the scattering
amplitude can be expressed in two equivalent ways,
either as the sum of resonances in the $s$ channel or
the particle exchanges in the $t$ channel.
However, it soon turned out, through various
developments in extending the formula, that the
requirement of the exact
$s$-$t$ duality was too restrictive for the theory of strong interactions.
The latter is then established
as the gauge field theory of `color' degrees of freedom,
{\it Quantum Chromo Dynamics}.

On the other hand, it also turned out that
there existed astonishingly beautiful and rich
structure behind the formula \cite{review}.  In particular, the basic 
dynamical
degrees of freedom governing the amplitude
was identified \cite{nambu} to be
 one-dimensionally extended objects, {\it strings}, instead of particles
in ordinary local field theories. One of the most important
aspects of the string picture was to make it possible to
identify  the symmetry ensuring the exact $s$-$t$ duality as
 the conformal symmetry in the
dynamics of strings.  Namely, the
world-sheets formed as the trajectory of
strings in space-time behave as Riemann surfaces.
Most of the restrictive features of the string theory
come from the requirement of the (super) conformal symmetry.

Here are some examples from the list of the
properties of strings which result from the
conformal symmetry:  The critical dimensions, namely
the conditions for the dimensionality
of space-times such as
26 in bosonic models and 10 in the models with fermionic
strings, are the conditions for
cancellation of the conformal anomaly which
usually plagues any quantum field theories to preserve the
conformal symmetry of classical theory.
The fixed-point equation in the renormalization-group
formulation of the world-sheet dynamics contains the
Einstein equations for background space-time
in the long-distance regime.  Cancellations of
the ultra-violet divergences associated with the
loop diagrams of string world sheets,
one-to-one correspondence between the
spectrum and the allowed external fields
for the propagation of strings,  .. and so on, are
also the direct consequences of the conformal symmetry.

All of these properties are crucial ingredients
which constitute the bases for the interpretation of string theory
as the unified quantum theory of all interactions including gravity.
 In particular, the string theory showed us, for the
first time, that there existed a natural avenue towards
the quantum theory of gravity
by circumventing \cite{shapiro} the non-renormalizability of the
ultra-violet divergences, since it turned out that
the spectrum of consistent string theories necessarily contain the graviton 
whose interaction in the
long-distance regime is correctly described \cite{yo}\cite{ss}
by the classical general relativity, adjoined with
some additional fields like dilaton, antisymmetric
tensor fields and further fermionic fields (called gravitinos) associated 
with space-time
supersymmetry \cite{gso}, namely, supergravity.

However, the present formulations of string theories
are far from satisfactory.  The string picture only
provides us a set of perturbative rules for constructing
the S-matrices in the expansion with respect to the
strength of the string coupling $g_s$ which is related to the
vacuum expectation value of the dilaton field $\phi$
as $g_s = e^{\phi}$.  That is, we have seen only the surface of
 string theory in the special asymptotic limit $\phi \rightarrow -\infty$.
  From experiences in gauge-field theories for unifying
the other interactions than gravity and the role of
the non-perturbative behaviors in QCD, we cannot
hope that such a perturbation expansion alone
is sufficient to explore the physics near Planck scale
where the very dynamics of the vacua and their
transitions must be investigated.
Present perturbative formulation has no power
in giving definite predictions in such regime. All what we could
envisage from the perturbative string theory
is that there must exist certain tremendous theoretical structure
whose real meaning and significance are not yet fully understood.

The developments \cite{pol1} occurred in recent few years are
very encouraging in this respect.
Firstly, various duality relations which connect
different perturbative string theories were discovered,
and passed many nontrivial checks.
The T-duality  relates different background
fields by the duality transformation on the world sheet
of strings.  For example, a string theory with
 a space-time dimension  which is compactified into a circle
with a radius $R$ is equivalent with a string theory
with the inverse compactified radius $\ell_s/R$, where $\ell_s$
is the length constant characterizing the space-time extension of strings 
\cite{ky}.  The S-duality \cite{sen}, on the other hand, interchanging the
roles of the conserved Ramond-Ramond
and  Neveu-Schwarz charges, relates a
string theory with a given string coupling $g_s$ to
a string theory with inverse string coupling $1/g_s$.
Furthermore, it turned out that, in addition to the
previously known perturbative string theories,
there must exist a new theory, called M-theory,
whose perturbative behavior in the long-distance limit is
described by 11 dimensional supergravity.
 From the viewpoint of M-theory \cite{witten1}, the type IIA string
perturbation theory is a reduced model in which one of the
spatial dimension is compactified with the radius
$R=g_s\ell_s$.  Hence, in the perturbative regime $g_s \sim 0$,
the effective space-time dimension is 10.

In all these developments, one of the most crucial
aspects is the discovery of the role of Dirichlet branes \cite{pol2}
(usually called `D-branes') which carry the
Ramond-Ramond charges.  The elementary string states
can never carry the Ramond-Ramond charges.
Thus D-branes made it possible to discuss the
properties of S-dual transformed theories in concrete forms.

Furthermore,  D-branes opened up entirely new
viewpoints for string theory.
It has been known that there are a large number of
classical solutions \cite{horostro} for the low-energy effective
field theories corresponding to the perturbative
string theories, which contain various types of
black-hole configurations.  From the viewpoint of
D-branes, most of them can be regarded as composite
systems of D-branes.  Once such identifications
are made, we can count the number of
physical states with given mass and charges,
at least in the limit of extremal black holes
where the masses of black holes approach to the
lower bounds proportional to the charges.
The latter limit corresponds to limits (called
the BPS limit) where the  space-time supersymmetry
is partially recovered
for the D-brane configurations.
The results of counting, obtained by neglecting
 the string interactions,  are reliable owing to the
protection of supersymmetry.
In this way, Bekenstein-Hawking's classical formula
for the black-hole entropy was  derived \cite{strovafa}
\cite{maldacena1},
for the first time,
by the quantum-statistical method.
In view of an analogy with the role played by
 black-body radiation in the
early development of quantum theory in the beginning of
this century,  the statistical derivation of the
black hole entropy should be regarded as an important
testing ground where we can check whether  or not
any theory of quantum gravity correctly identifies
the basic microscopic degrees of freedom.
Thus string theory now turned out to be really the
qualified candidate for the theory of quantum gravity
from the viewpoint of black hole physics too.
Not only that, these achievements clearly indicate that
the space-time structure near Planck length or shorter
should be interpreted in a drastically different way from the
classical geometrical picture, as has been emphasized in
many occasions from various different perspectives.

Another important new aspect of string theory
suggested from the advent of D-branes is the possibility
of reconstructing string theory
by using the D-branes as fundamental
degrees of freedom.  In other words, the
original string may now be interpreted as a sort of
collective modes in such a formulation.
The hope in doing this is that such a reformulation
may well be suitable to explore the non-perturbative
regime of string theory and, more ambitiously,
to obtain the rigorous non-perturbative definition
of the theory.

In the usual perturbative string theory,
the D-branes are treated by introducing open
strings with Dirichlet boundary conditions (hence the
name `Dirichlet' brane), which impose 
the positions of the D-branes  to be the ends of the
open strings.  Then the positions of the D-branes
can be regarded as collective coordinates
of string theory.  The low-energy effective theory
for such collective coordinates \cite{witten2} are
the super Yang-Mills theory, whose interpretation is of course
entirely different from the usual one as for the
unified theory of forces other than the gravity.
Namely, the theory lives on the world volume
of the D-branes, and the diagonal parts of the
Higgs fields now play the role of transverse coordinates
of the D-branes.  All of other degrees of freedom,
i.e. off-diagonal components and the gauge fields,
are supposed to describe the dynamics of short open strings
interacting with D-branes through the Dirichlet
condition.  Attempts towards the reconstruction of string theory
are based on these effective super Yang-Mills theories.
The first attempt along this line,
so-called Matrix theory \cite{bfss}, is to hypothesize
the lowest dimensional D-brane in the type IIA
theory, zero-dimensional branes=D-particles,
as the fundamental building blocks, or partons,
 of the theory and
the effective super Yang-Mills theory with 9
Higgs fields, i.e. 9 dimensions,  to be
the exact theory in the infinite momentum frame (IFM)
in 11 dimensional space-time.
Another attempt called the type IIB matrix model \cite{ikkt},
followed after this proposal, is to
identity the zero-dimensional super Yang-Mills theory with
10 Higgs fields,
namely, just the dimensional reduction of the
10D super Yang-Mills to a point, as the
underlying non-perturbative theory for the type IIB strings
in 10 dimensional space-time.  In both cases,
taking the large $N$ limit is assumed to be an
essential requirement for the exact nonperturbative
formulation of the theory.  In the former case,
there is some indication that the theory
is meaningful even for finite $N$ in the viewpoint
of the discrete light-cone quantization \cite{suss2}\cite{seiberg}.
Although there is no definitive verification
that these proposals can be accepted as the correct
definitions of the nonperturbative string theory or
M theory,
we already have reasonable amount of evidence
for believing that these are indeed promising starting
points, at the least,  towards our goal.

Besides various technical problems, one of the
mysteries associated with these proposals is what
the underlying principle which selects the
super Yang-Mills theory is, and why such Yang-Mills theory,
provided it is the correct starting point,
could give rise to  the consistent theory of
all interactions including gravity.
For example, we do not know definite symmetry structure
in super Yang-Mills theory which ensures the emergence of General Relativity 
in the long-distance regime, except for the
existence of maximal supersymmetry which is
only global from the viewpoint of the target space-time.
It is not at all clear whether the maximal supersymmetry
is sufficient to guarantee
the emergence of consistent (namely,
finite and unitary) quantum theory of gravity,
beyond its role in effective low-energy descriptions
of near BPS dynamics of D-branes.
For example,  we do not know any
systematic method for extending
the matrix models to general
backgrounds, since we do not have a firm
understanding on what the degrees of freedom
in the super Yang-Mills descriptions representing the
backgrounds are. In the case of Matrix theory, in particular,
how to formulate the theory in Lorentz invariant
fashion remains as a big mystery.  In the case of
the type IIB matrix model, although Lorentz covariant
by its definition, we do not know how to discuss
the geometry of backgrounds fields even in the
long-distance approximation,
and hence we cannot discuss the symmetry
under the S-duality transformation which
is expected to be the exact symmetry of the theory.
These difficulties
clearly indicate that some crucial things are missing
in our present understanding of the matrix model
approaches towards non-perturbative string or M theory.

 Thus it seems to us that one of the most important
tasks now is to seek for the definite guiding principles
which explain the emergence of Yang-Mills theory and
on the bases of which we can proceed
toward more satisfactory formulation
and definition of the theory.
In the following, we review some of the
attempts along this line of thoughts from two
specific viewpoints, namely, the
space-time uncertainty relation as a possible
universal characterization of the short-distance
space-time structure in string theory, and the possibility of
microscopic theory of black holes from the
viewpoint of D-brane matrix models.
The reason why we emphasize these two
viewpoints must now be clear. Compared with
classical General Relativity and gauge-quantum
field theories of other interactions than gravity,
the unified theory should be characterized by
its departure from the former with respect to
the short-distance space-time structure and
the fundamental microscopic degrees of freedom
at the shortest possible space-time scale.

In section 2, we motivate and
summarize the space-time uncertainty
principle \cite{yo2} and discuss its roles in the dynamics of
D-branes, especially in the case of D-particles.
In section 3, we briefly review the entropy problem of
black holes and discuss the statistical derivation
of entropy from Matrix-theory viewpoint.
In section 4, we will suggests a generalized
space-time uncertainty relation suitable for
M theory and conclude
by pointing out that  the space-time uncertainly
relation expains some of the crucial 
properties leading to the holographic bound 
for the entropy of D3 brane. 

\section{Space-time uncertainty principle of string theory}
\subsection{World-sheet conformal symmetry
and the short-distance space-time structure}

We have emphasized that the symmetry
which leads to the key features of string theory as
the unified theory of all interactions is the
conformal symmetry exhibited in the dynamics of
world sheets of strings. It is therefore reasonable to
expect that the conformal symmetry
must be taken into account in
the nonperturbative reformulation of the theory.
However, the conformal symmetry as the
symmetry of the world-sheet dynamics seems to be
intrinsically a perturbative concept,
depending  upon the emergence of
Riemann surfaces in the perturbative formulation.
The Riemann surfaces in string theory ,
or more precisely, their
equivalence classes called the
moduli space of Riemann surfaces,
replace the role of the space of proper times
for relativistic point particles.
The proper time is intrinsically a
perturbative concept in quantum field theory.
It is therefore important
to reinterpret the conformal symmetry
using physical quantities which do not
intrinsically depend upon the perturbation theory.

The importance of reinterpreting the
conformal symmetry is also expected from
the following consideration.
The world-sheet dynamics of perturbative
string theory is constructed using the conformal
field theories (CFTs).  From the general viewpoint of
two-dimensional quantum field theories,
CFTs are a very special class of field theories,
characterized by the fixed-points of
renormalization-group equation.  In other words,
the astonishingly desirable features and consistency
of string theory as the unified theory are realized only by
choosing a special class of two-dimensional field theories
describing the dynamics of world sheets.
We can point out an amusing analogy here.
The requirement of conformal symmetry is
playing a somewhat analogous role with the Bohr-Sommerfeld
quantization condition for the
adiabatic invariants in classical analytical dynamics.
The consistency of imposing the latter condition
in classical theory
comes from the adiabatic theorem which says that
the adiabatic invariants remain invariant under
slow changes of external variables. Hence
the quantization condition on them, although ad hoc
from the viewpoint of the continuous degrees of freedom of
possible initial conditions, can be protected against
small perturbations.
The analog to the adiabatic theorem in quantum
field theories is the fixed-point condition
of renormalization group.  Of course, by the advent of
quantum mechanics, the ad hoc quantization
conditions were replaced by the operator algebra
of physical observables which are characterized by
canonical commutation relations.
Similarly, we should perhaps
 expect that the requirement of conformal
symmetry in the perturbative formulation of string theory
is replaced by some more general
condition which can be expressed without using the
intrinsic properties of rules for constructing string
S-matrices.

  With these motivations in mind, one of the present authors
proposed to reformulate the world-sheet conformal
symmetry as a simple uncertainty relation in
space-time \cite{yo3}. Let us first briefly recapitulate the
space-time uncertainty relation.
\footnote{
Independently of
the proposal of the
space-time uncertainty relation,
an intimately related
concept of the `minimal length'
was suggested by several authors
\cite{ven}\cite{gross} and has got
popularity \cite{review3}.  It should be noted that
the space-time uncertainty relation is more
general than the concept of the minimal
length which now turned out to be
restricted to the perturbative string amplitudes.
}

Consider a parallelogram on the string world-sheet  and the
Polyakov amplitude for the mapping from the
world-sheet to a region of the target  space-time.
Let the lengths of the two orthogonal sides in the
world-sheet parallelogram be
$a, b$ in the conformal gauge
where $\dot{x} \cdot x'=0, \,
\dot{x}^2 +x'^2 =0$  and the
corresponding physical
space-time length be $A, B$, respectively.
Then apart from the
power behaved pre-factor, the amplitude
is proportional to
\EQ
\exp [-{1\over \ell_s^2} ({A^2\over \Gamma}
+{B^2 \over \Gamma^*})]
\label{abamp}
\EN
where
\EQ
\Gamma \equiv {a\over b}, \, \, \Gamma^* \equiv {b\over a} ,
\quad (\Gamma \Gamma^* =1).
\EN
Due to the conformal invariance, the amplitude
depends on the Riemann sheet parameters only through
the ratio $\Gamma$ or $\Gamma^*$,
which are called the {\it extremal length} and the {\it conjugate
extremal length}, respectively.  Clearly, the relation
$\Gamma \Gamma^*=1$ leads to the uncertainty relation \cite{yo3}\cite{yo4}
\EQ
\Delta T \Delta X
 \sim \ell_s^2 \sim \alpha'
\label{spacetimeuncertainty}
\EN
taking the $A$ direction to be time-like
$\Delta T\sim \langle  A \rangle$ and
hence the $B$ direction to be space-like
$\Delta X \sim  \langle  B\rangle$.
The obvious relation $\Gamma \Gamma^*=1$ is the
origin of the familiar modular invariance of the
torus amplitudes in string theory.  The extremal
length is the most fundamental moduli
parameter characterizing conformal invariants,
in general.  Since arbitrary
amplitudes can be constructed by
pasting together many parallelograms, any string
amplitudes satisfy the above reciprocal relation
qualitatively.  Although this form looks
too simple as the characterization of the
conformal invariance, it has a virtue
that its validity  is very general, as we
will explain shortly,  and does not use the
concepts which depend intrinsically  on
perturbation theory.  Our proposal \cite{yo4}\cite{liyo} is to
use this relation as one of possible guiding
principles towards nonperturbative
reformulation of string theory and M-theory.

We note that the above uncertainty relation
suggests that the classical space-time is
something very analogous to the classical
phase space of particle mechanics.
Indeed the form (\ref{abamp}) of the amplitude
is reminiscent of the Wigner representation of the
Gaussian wave packet of a point particle.
In this sense, our proposal is similar 
to other proposals \cite{other} of space-time quantization.
Most of such proposals, however, presuppose the
existence of minimum length in all
space-time directions.  Our proposal is
quite different in this respect, since
the above relation does not prohibit the
existence of shorter lengths than the string scale
$\ell_s$ in a particular direction. It only
says that there is a duality  between short and
large distance phenomena in string theory.
The importance of shorter length scales from
the more recent context of various duality relations
in string theory has been emphasized by Shenker in ref.
\cite{shenker}.

The uncertainty relation (\ref{spacetimeuncertainty})
is consistent with an
elementary property of strings that the energy of
a string is roughly proportional to its space-time length.
\EQ
\Delta E \sim  {\hbar \over \alpha'}\Delta X_l .
\EN
with $X_l$ being the length of a string measured
along its longitudinal direction.
Then the ordinary time-energy uncertainty relation
$\Delta T \Delta E \ge \hbar$ leads to (\ref{spacetimeuncertainty}) \cite{yo3}.
It is important here to discriminate the length scales in the longitudinal
and transverse directions with respect to the string.
  As is well known, the (squared)
transverse length scale grows  logarithmically with energy used to probe
the strings.
This explains the linearly rising Regge trajectory
for the Regge-pole behavior in  high-energy peripheral
scattering.
\footnote{Note that in the special case of the strings in
1+1 dimensions, there is a longitudinal size but no transverse size,
 and, hence,  the
high-energy limits of the perturbative amplitudes show  power-law  
behaviors,
apart from the phase factors associated with the
external legs.  For a discussion on the physics of the 1+1 dimensional
string scattering, we refer the reader to ref. \cite{jeliyo}.
For a review on the connection of 1+1 dimensional
strings and old-fashioned matrix models, see
\cite{yokikk} and references therein. }
The dual role of the time and the longitudinal
spatial lengths
is a natural space-time expression of the original
$s$-$t$ duality.
The particle exchange (Regge exchange
in the old language)  and
the resonance behaviors correspond to the
regimes, $\Delta X_l \rightarrow \infty$ and
$\Delta T \rightarrow \infty$, respectively.
Furthermore, the Regge behavior is
consistent with the existence of graviton,
since scattering amplitudes in general
are  expected to be roughly proportional to $\Delta X_l \sim \ell_s^2 
/\Delta T
\propto E$ which
implies, by adopting the argument
in \cite{susskind2},  that the intercept $\alpha(0)$ of the leading Regge 
trajectory
is 2, from the relation $E \sim E^{\alpha(t)-1} $.

On the other hand,
in the high energy fixed-angle scatterings with
large $s$-and $t$-channel momenta studied
in detail in \cite{gross},
we are trying to probe the region where both the time and the
spatial scales are small. Clearly, such a region is incompatible
with the space-time uncertainty relation. The exponential fall-off
of the perturbative string amplitudes in this limit
may be interpreted as a manifestation of this property.
According to the space-time uncertainty relation, at least
one of the two length scales, $\Delta X$ or
$\Delta T$,  must be larger than the
string scale $\ell_s$.  Therefore there is no degrees of
freedom left in this regime.
It is well known that any consistent theory of quantum gravity
must indicate lessening of the degrees of freedom
near the Planck scale where the quantum nature of gravity
becomes important.
We have seen that the space-time uncertainty relation
can indeed be a natural mechanism for this.
Qualitatively, it is also consistent with the
known high-temperature behavior \cite{attickwitten} of the
perturbative string amplitude, since the
high-temperature limit is effectively
equivalent to considering the limit $\Delta T\rightarrow 0$.

\subsection{Black hole physics and the space-time
uncertainty relation}
The space-time uncertainty relation is also
expected to be important in black hole physics,
since black hole provides us a natural
tool  for probing the
short time region.  By definition,
the black hole horizon is the region where,
owing to infinite red shift effect,
any finite time interval in the asymptotic space-time
boundary corresponds to infinitely short time interval
$\Delta T\sim 0 \rightarrow \Delta X\rightarrow \infty$.
Thus we expect that the black hole scatterings can be
 a unique arena where we may check
the stringy behavior of the theory.
Perhaps the black hole horizon is
one of the places where the predictions and interpretations
of string theory could significantly deviate
from those of local field theory, since
there we cannot rely upon the usual
intuition based on familiar local field theory.
For example, some peculiar behaviors
found for a matrix model for the
1+1 dimensional black hole \cite{defmat} might be
due to this. For a review on this controversial issue,
we refer the
reader to \cite{yo-kikk} and the references therein.
Unfortunately,  most of the
reliable discussions of the black-hole
scattering of strings are based on the
low-energy limit ($\alpha' \sim 0$).
However, the recent developments of the
D-brane technology and Matrix theory
indeed suggest
a picture of the black-hole scattering which is
quite different from the local field theory with respect
to the puzzle of information loss \cite{susskind3}.

Qualitatively, the deviation from the
local field theory and string theory
should manifest itself when the length scale
of black holes becomes comparable to the
string scale where the space-time uncertainty
relation becomes essential. There is indeed a signature
that the stringy nature becomes
dominant beyond this regime.
In D-dimensional space-time, the
Bekenstein-Hawking entropy
of a Schwarzschild black hole of mass $E$ is
given by
\EQ
S  \, \, (= {{\rm Area}\over 4G_N})=
{1\over 4}G_N^{{1\over D-3}}E^{{D-2\over D-3}} .
\EN
On the other hand, the massive (super) string state
has a quantum degeneracy of the order
\EQ
d(E) = \exp 2\pi \sqrt{{D-2\over 6}N}
\EN
with $N= (\ell_s E)^2$. If we identify the
Schwarzschild black hole of string size
$\ell_s$ with a massive string state by
setting \footnote
{This argument was given,
 independently of ref. \cite{horopol},  in a lecture
at the 96 Japanese  GRG workshop  by one of the
present author \cite{yoneya-grg} .
}
 \[
\ell_s \leadsto \ell_{{\rm eff}} \sim R_{{\rm schwarzshild}} , \sim (G_N 
E)^{{1\over D-3}} ,
\]
we have
\EQ
\log d(N) \sim G_N^{{1\over D-3}} E^{{D-2\over D-3}} \sim S_{{\rm BH}} .
\EN
This strongly
 suggests that the black hole with sizes
comparable with string scale must be treated by
string theory, and the latter would possibly provide
the microscopic explanation of the entropy.
More detailed examination of the correspondence
between string states and Schwarzschild black holes
was studied by Horowitz and Polchinski in ref. \cite{horopol}.
Further discussion on the entropy problem from the
viewpoint of Matrix theory and the relation between the
space-time uncertainty relation and
black-hole physics is the subject of later sections.

\subsection{The space-time uncertainty relation for
D-particles}

One of the puzzles in proposing the space-time uncertainty
relation is how to probe the region $\Delta X\sim 0$
using strings. Since this leads to large times
$\Delta T\rightarrow \infty$,
the original interpretation was that it provides an
explanation why the asymptotic states of any string states,
in spite of their intrinsic spatial extension,
can be treated as the {\it local} space-time fields,
as in the nonlinear $\sigma$ model approach
to the computation of string S-matrices.
Because of this property,  the vertex
operator of graviton  is exactly
given by  the space-time energy-momentum tensor
of strings and hence ensures the equivalence
principle.  It is, however, desirable to probe the
region of short spatial scales directly.

The D-branes, in particular, the point-like D-brane
(D-particle), 
turn out to be ideal probes for that purpose.
A D-particle in type IIA superstring theory is
characterized by its Ramond-Ramond charge
associated with the 1-form RR gauge
field, which is identified with the
Kaluza-Klein momentum in the 11th direction
with compactified radius $R\sim g_s\ell_s$.
  From 10 dimensional viewpoint, it simply behaves
as a point particle with mass $1/g_s\ell_s$,
interacting with each other and with closed strings
 through open strings
whose ends are attached to D-particles.
If we apply the space-time uncertainty relation to
the open strings  between D-particles,
we expect that the uncertainty relation holds
between the time scale $\Delta T$ and the
spatial scale $\Delta X$ which is 
transverse to  D-particles.  As shown in \cite{liyo},
we can readily derive the
characteristic spatial and temporal scales using this relation.
Let  us repeat the simple argument here.

Consider the scattering of two D-particles of mass $1/g_s\ell_s$ with
the impact parameter of order $\Delta X$ and the relative
velocity $v$ which is assumed to be much smaller than the
light velocity. Then the characteristic interaction time $\Delta T$ is
of order ${\Delta X\over v}$. Since the impact parameter is of the
same order as the longitudinal length of the open strings
mediating the interaction of the D-particles, we can
use the space-time uncertainty relation in the form
\[
\Delta T \Delta X \sim \ell_s^2 \Rightarrow
{(\Delta X)^2 \over v} \sim \ell_s^2
\]
This gives the order of the magnitude for the  minimum possible
distances probed by the D-particle scatterings with
velocity $v \, (\ll 1)$.
\EQ
\Delta X\sim
\sqrt{v}\ell_s.
\label{velocity-distance}
\EN
To probe short spatial
distances, we have to use D-particles with small
velocity.  However, the slower the
velocity is and hence the longer  the interaction time is,
the larger is the spreading of the wave packet.
\EQ
\Delta X_w \sim
\Delta T\Delta_w v \sim
{g_s\over v}\ell_s
\EN
since the ordinary time-energy uncertainty relation
says that the uncertainly of the velocity is of
order $\Delta_w v\sim g_sv^{-1/2}$ for the time
interval of order $\Delta T \sim v^{-1/2}\ell_s$.
Combining these two conditions, we see that the
shortest spatial length is given by
\EQ
\Delta X \sim  g_s^{1/3}\ell_s
\label{plancklength}
\EN
and the associated time scale is
\EQ
\Delta T \sim  g_s^{-1/3}\ell_s .
\EN
(\ref{plancklength}) is of course the
11 dimensional Planck length which
is the characteristic length of M-theory which was
first derived in the super YM context in \cite{dkps}.
As argued in ref. \cite{liyo}, it is actually possible to
probe shorter lengths than the Planck length
if we consider a D-particle in the
presence of many (=$N$) coincident D4-branes.

Above derivation of the Planck length
emphasizes its meaning as the minimal
{\it observable} distance in a dynamic process
of D-particles. We can also derive the
same length scale from a slightly different
viewpoint of the distance scale below which
we cannot discriminate the quantum state of
D-particles.
For a given velocity $v$ of a D-particle,
the  minimal possible distance among D-particles is
\EQ
\Delta X \ge \sqrt{v}\ell_s .
\label{eq2}
\EN
The ordinary Heisenberg
relation $\Delta p \Delta X \ge 1$, on the other hand,
gives a lower bound for the velocity
 \EQ
v \ge  {g_s\ell_s \over \Delta X}
\label{eq3}
\EN
which, by combining with (\ref{eq2}),
 leads again to the minimum distance
(\ref{plancklength}).  Thus, the space-time uncertainty
relation clearly says that for each state of a D-particle
no information can be stored below the Planck
distance and the minimum bit of information
carried by a D-particle is given by the Planck volume
$V_P \sim \ell_P^9$
in the transverse space.  This is certainly consistent with the
holography principle \cite{thooft-suss} in M theory interpretation of 
D-particles;
however, it does {\it not} yet
prove that the latter is a direct consequence of
the space-time uncertainty principle, since
we have to actually deal with many-body
dynamics of D-particles.  Further discussion
on the possible relation of the space-time uncertainty
relation and the holography will be given in
section 4 after summarizing, in section 3,  the recent
developments in black hole physics from the
viewpoint of Matrix theory.

\subsection{The scale  symmetry
of the effective Super Yang-Mills theory and the
space-time uncertainty relation}
Let us now see how the space-time uncertainty
principle is encoded in the effective (super)
Yang-Mills theory of D-branes. For definiteness,
we first concentrate to the case of D-particle and
then indicate the generalization to general D-branes
only briefly.

The collective variables of $N$ D-particles are the
 one-dimensional $N\times N$
Hermitian matrix field  $X_i(t)$
and their fermionic partners,
where $t$ is the time along the world line of the
D-particle.
 \EQ
S =\int dt \, \Tr
\Bigl( {1\over 2g_s\ell_s} D_t X_i D_t X_i + i \theta D_t \theta
+{1 \over 4g_s\ell_s^5} [X_i, X_j]^2 -
{1\over \ell_s^2}\theta \Gamma_i [\theta, X_i]\Bigr)
\label{d0action}
\EN
where the covariant derivative is defined by
$D_t X ={\partial \over \partial t}X +[A, X]$
with $A$ is the U($N$) gauge field.
This action is invariant under the scale
transformation
\EQ
X_i(t) \rightarrow X_i'(t') =\lambda X_i(t) ,
\quad
A(t) \rightarrow A'(t')=\lambda A(t), \quad t\rightarrow t'=\lambda^{-1}t
\label{spacetimescale}
\EN
\EQ
g_s \rightarrow g_s'=\lambda^3 g_s .
\label{stringcouplingscale}
\EN
Note the opposite scalings for transverse
coordinate $X_i$ and the time in
(\ref{spacetimescale}). The
scaling of the coupling constant (\ref{stringcouplingscale})
leads to the characteristic spatial and
temporal scales given before on the basis of the
space-time uncertainty relation.  Thus the
effective super Yang-Mills theory naturally
encodes the space-time uncertainty principle.
This scaling property can be readily extended to
general D$p$-brane:
\EQ
g_s \rightarrow g_s'=\lambda^{3-p} g_s
\EN
with the same transformation for the transverse
coordinate and time as (\ref{spacetimescale}).
The uncertainty relation for D-branes in general  says that
the long distance phenomena in the (transverse) target space
is dual to the short distance phenomena in the world volume and {\it vice 
versa}.
In particular, for D3-brane ($p=3$) where the
string coupling is invariant, the Yang-Mills theory
is invariant corresponding to the scale invariance
of the 3+1 dimensional Yang-Mills theory.
This means that the dynamics of D3-brane involves
all the scales in both the target and world volume,
keeping the dual nature of them.  For other ``dilatonic" branes,
on the other hand,
the effective scales are fixed by the vacuum expectation values
of the dilaton.  We emphasize that the space-time
uncertainty relation itself is valid for general D-branes, irrespectively of such specialities.

In the new interpretation of super Yang-Mills
theory as the effective collective
field theory of D-branes, the
participation of the string coupling
in the scale symmetry is allowed, if
we remember that the string coupling constant $g_s$  can  in principle be
treated as a dynamical variable corresponding
to the vacuum expectation value of the dilaton.
Although in the present formulation of the
strings and D-branes, the string coupling appears
as an external parameter, we should expect that
the string coupling should be eliminated from
the fundamental (hopefully background independent) formulation of the 
theory. For example, in the
string field formalism, we can indeed eliminate the
string coupling by making a shift for the
string field as shown in \cite{yo5}.

Now the scaling property constrains  the effective action
for D-particles with respect to a double power series
expansion in $1/r$ (inverse distances) and $v$
(velocities), if we neglect the
second and higher time derivatives, as
\EQ
S^{eff} =\int dt \sum_{L=0}^{\infty} \sum_{A+2B=L,  \, B\ge 0} 
R^{3L-3+A+B}\ell_P^{-6L+6-3(A+B)}
\times r^A v^B c_{A,B}
\EN
where
\footnote{
This form is first pointed out  in ref. \cite{bbpt}.}
 $R=g_s\ell_s, \ell_P =g_s^{1/3}\ell_s$ are
the compactification radius and the Planck
length of M-theory, respectively, and $L$ denotes the number
of loops. Here, $r$ and $v$ collectively represent
the distances and velocities, respectively, of
many D-particle systems.  The numerical
coefficients $c_{A, B}$ depends on the
number $N$ of D-particles as
\[
c_{A, B} =N^{L+1}\sum_{0\le n \le L/2} c_{A,B,n}N^{-2n}.
\]
In terms of the string length and the string
coupling, the expansion is a double
expansion
\EQ
{1\over g_s\ell_s }({g_s\ell_s \over r^3})^{L-4/3} ({\ell_s^2 v \over 
r^2})^B .
\EN
Note that the characteristic combination
$\ell_s^2 v/r^2$ just reflects the space-time
uncertainty relation (\ref{velocity-distance})
for the velocity and the distances, while the
the combination $g_s\ell_s /r^3$ reflects the
characteristic spatial scale (\ref{plancklength}).
Thus the space-time uncertainty relation
naturally explains the emergence of
 the two characteristic quantities
by which we can measure the effect of string
extension and string coupling in the effective
super Yang-Mills model.  Of course, the
power series expansion loses its meaning when
these quantities becomes of order one.
We mention that the emergence of the
characteristic spatial scale (\ref{plancklength})
is also observed \cite{giddings}
in a different version of Matrix theory, the so-called
Matrix string theory \cite{matrixstring}.

We also note that in terms of the 11 dimensional
Newton constant $G_{11}= g_s^3\ell_s^9  $,
these quantities become ,with $B=2n$,
\[
({\ell_s^2 v \over r^2})^B \rightarrow
{G_{11}^{2n/3}\over R^{2n}}
({v^2 \over r^4})^n ,\quad
({g_s^{1/3}\ell_s \over r})^{3L} \rightarrow G_{11}^{L/3}r^{-3L} .
\]
The classical supergravity
contribution corresponds to the
"diagonal" terms $L+1=n$,
$v^{2(L+1)}/r^{7L}$.  It seems reasonable to
interpret the other terms with $L > n-1$ or
$n > L+1$ as the quantum gravity and
stringy corrections, respectively.
The diagonal contributions have been shown to
precisely agree with classical
11 dimensional supergravity up to
two loop approximation for
2 \cite{bb}\cite{bbpt} and 3 body
\cite{okyo} interactions
\footnote{
There has been some controversy in the case of
3-body interactions, since the first paper in ref. \cite{dine}
on this. Ref. \cite{okyo}  however showed that the
precise agreement holds if we take into account
all the relevant terms both in 11D supergravity and
Matrix theory, except the recoil effect.
Since in this classical approximation, the effective
actions in both sides are identical, the recoil effect should
also agree.
} of D-particles for finite $N$.
Supersymmetry shows that there is no correction
for  $n=0, 1$ terms to the classical contributions.
Explicit computations in these references further showed that
the correction to the $n=2$ term also vanish up to the two-loop
approximations, confirming the supersymmetric
non-renormalization theorem (see, e. g., \cite{paban}).

\subsection{The Maldacena  conjecture and
the space-time uncertainty principle}

If the above scaling property is interpreted as
the symmetry of the theory, it is natural to ask
whether the scaling symmetry can be generalized to
full conformal symmetry, since we know that
the scaling symmetry of the D3-brane is
embedded in the full (super) conformal symmetry
of the theory.  Because of the existence
of the maximal supersymmetry, we expect that the
full conformal symmetry is unbroken and hence
puts stronger constraints on the dynamics of
D-particles than the simple scaling
arguments. In a recent work \cite{jeyo}, it was shown that
there exists such a generalization.
The D-particle action is indeed invariant under the
infinitesimal coordinate transformation
\EQ
\delta_K X_i = 2  t X_i , \, \, \delta_K A= 2  t A , \, \,
 \delta_K t =-  t^2 , \, \, \delta_K g_s =6  t g_s .
\label{d0specialconf}
\EN
 With the scaling symmetry $\delta_D$ and the trivial time
translation symmetry $\delta_H$,  we have the
1+1 dimensional conformal algebra
\EQ
[\delta_D, \delta_H] = \delta_H , \, \,
[\delta_D, \delta_K] =-\delta_K , \, \,
[\delta_H, \delta_K]=2\delta_D  .
\label{su11confalg}
\EN

This is important in view of the conjecture
made by Maldacena \cite{maldacena2},
who proposed essentially that the
correspondence between the
super Yang-Mills theory and the supergravity
implied by the nature of the former as the
effective description of the string theory becomes
exact in the limit where the higher-excitation modes
of open strings are decoupled in the former and the
strong curvature and quantum gravity effects are
 neglected in the latter.  The first condition
leads us to consider the near horizon region of the
D-branes, while the second condition requires the
large $N$ limit with fixed but large $g_sN\sim g_{{\rm YM}}^2 N$.  In the 
particular case of D3-brane,
the near horizon geometry is described by the
Anti de Sitter geometry in 4+1 dimensions, $AdS_5$.
It was then proposed \cite{klegbupol}\cite{witten-holo} to interpret
the 3+1 dimensional super Yang-Mills theory
as the boundary conformal field theory
corresponding to type IIB supergravity in
the background of the conformal symmetric  $AdS_5$.

In ref.  \cite{jeyo}, it was shown that there exists a similar
conformal symmetry for the background fields
of D-particle.  For example, the 10 dimensional
(string frame) metric of $N$ coincident D-particles are
given by
\EQ
ds_{10}^2 = -e^{-2\tilde{\phi}/3}dt^2 + e^{2\tilde{\phi}/3}dx_i^2 .
\label{d0metric}
\EN
where
\EQ
e^{\phi}= g_s e^{\tilde{\phi}}, \, \, \,
e^{\tilde{\phi}} =\bigl( 1 +{q\over r^7}\bigr)^{3/4}
\EN
with $q=60\pi^3 (\alpha')^{7/2} g_sN$. If we take the
decoupling limit $\alpha'\rightarrow 0$ keeping
\EQ
U \equiv  {r \over \alpha'} ,
\quad  g_{YM}^2 \equiv  {g_s\over \alpha'^{3/2}}
\EN
fixed, the metric is reduced to the form
\EQ
ds_{10}^2 =\alpha'\Bigl(-
{U^{7/2}\over \sqrt{Q}}dt^2
+{\sqrt{Q}\over U^{7/2}}
\bigl(dU^2 + U^2 d\Omega_8^2\bigr)\Bigr)  .
\label{decouplinglimit}
\EN
It is easy to check that the metric and the dilaton is
invariant under the same scale transformation
as before
\EQ
U\rightarrow \lambda U ,
\, \,
t \rightarrow \lambda^{-1}t ,
, \,
g_s \rightarrow \lambda^3 g_s ,
\EN
Furthermore, they are also invariant under the
infinitesimal transformation
\EQ
\delta_K t = -  (t^2 +k{g_{YM}^2\over U^5}) ,
\label{eq315}
\EN
\EQ
\delta_K U =2  tU ,
\label{eq316}
\EN
\EQ
\delta_K g_s=6  t g_s ,
\label{eq317}
\EN
where $k$ is a constant independent of the string coupling
$k=24\pi^3 N$.  Together with the scale and time translation
symmetry, they form the same conformal algebra as
for the super Yang-Mills quantum mechanics.
The appearance of the {\it field dependent}
term, second term in (\ref{eq316}),  is similar to the case
of D3 brane. It is now very natural to suppose the
validity of the Maldacena conjecture for D-particle case,
as in the same sense as for the D3-brane. Namely,
the effective super YM quantum mechanics may be
interpreted as
the  {\it boundary} quantum theory of the
D0 metric. For more details
on this we refer the reader to \cite{jeyo}.
Here, however, we want to emphasize that there is
an unsolved question in Maldacena conjecture in general.
The question is how to derive the field dependent
transformation laws in the bulk from the linear
transformation laws in the boundary field theory.

It should be noted that the
scaling symmetry naturally leads to the structure
which is consistent with the
space-time uncertainty relation $\Delta U \Delta t \sim 1$
for the coordinates $U$ and $t$ just as
$\Delta X \Delta T \sim \alpha'$ for the
small proper distance $
\Delta X \sim \sqrt{\alpha ' Q^{1/2}U^{-7/2}}
\Delta U$ and proper time $\Delta T \sim \sqrt{\alpha ' U^{7/2}Q^{-1/2}}
\Delta t$, the latter of which vanishes, 
for fixed coordinate uncertainties,  in the
near core limit $U \rightarrow 0$ 
while for large $U$ the situation becomes opposite
$\Delta X \rightarrow 0$.
We also want to emphasize that the Maldacena's conjecture
can be regarded as a special case of the original
$s$-$t$ duality between closed and open strings.
Therefore,  its
consistency with the space-time uncertainty relation
and the appearance of the conformal symmetry
here again are not at all accident.

It seems natural to interpret the conformal symmetry
structure as a mathematical realization of the space-time
uncertainty principle, being the transformation groups
which leave the form of the super Yang-Mills action
invariant. The maximal supersymmetry is crucial
here in order to preserve
 the classical scale and conformal structure
in quantum theory.   The conformal symmetry
in the form (\ref{eq315}) is
then very powerful, since it
uniquely determines, combined with the
supersymmetric non-renormalization theorem,  the classical
contribution, i.e. the $L=n+1$ terms, of the effective action
to all orders in the velocity expansion as
shown in \cite{jeyo} .
The resulting effective action precisely coincides with the
effective action obtained \cite{bbpt} from the so-called
discrete light-cone interpretation of the
11 dimensional supergravity. Hence it
supports the interpretation of the
effective super Yang-Mills theory of
D-particles as the non-perturbative
formulation of M theory in the discrete
light-cone quantization scheme.
Not only that, our arguments almost
indicate that the most essential dynamics of D-particles
are encapsulated in the space-time uncertainty
principle
 as expressed in the form of the
generalized conformal symmetry.
Unfortunately, however, our
discussion is yet confined to a very special
background configurations of D-particles.
Therefore, one of the most important
direction from this viewpoint would be
 to generalize the arguments to arbitrary
backgrounds.

These structure is not special for D-particle and
D3-brane. At least formally, we can extend the
conformal symmetry and its consequences to arbitrary $p$
as will be discussed in \cite{jekayo}.

\subsection{Type IIB matrix model and the
space-time uncertainty principle}

The D-particle is the lowest dimensional objects in
type IIA string theory and M-theory. On the other hand,
in type IIB theory, the objects of lowest dimensionality are
D-instantons, namely $p=-1$ D-brane.
For the D-instanton, all of the
space-time directions, including even the time direction,
becomes the transverse directions.
The interaction of the D-instantons are therefore
described by the open strings which obey the
Dirichlet boundary condition with respect to all the
space-time directions. Hence the effective action for
D-instantons, at least for
low-energy phenomena, are described by the $10$ dimensional
super Yang-Mills theory reduced to a space-time point:
\EQ
S_{{\rm eff}}= -{1\over g_s}\Tr ([X_{\mu}, X_{\nu}])^2 +
{\rm fermionic
\, \, \, part}
\EN
which is covariant from the beginning.  We here suppress
the string scale by taking the unit $\ell_s=1$.

The dynamics of D-instantons is very special in the sense that
the only ``physical" degrees of freedom  are the lowest massless modes
of the open strings. The reason is that only
on-shell states satisfying the Virasoro condition
are the zero-momentum ``discrete" states
of the Yang-Mills gauge fields which are
identified with $X_{\mu}$'s and their
super partners.
The infinite number of the
 massive degrees of freedom of the open strings
should be regarded as a sort of
auxiliary fields.
We then expect that the
exact effective action for them should be
expressible solely in terms of the collective fields which
are nothing but the lowest modes $X_{\mu}$.
If we further assume that all the higher dimensional
objects including the fundamental strings could be various
kinds of  bound states or collective
modes composed of infinitely many
D-instantons,
we may proceed to postulate that
a matrix model defined at a single point
might give a possible nonperturbative
definition of the type IIB string theory.

This is a very bold assumption:
For example, unlike the case of the M(atrix) theory,
here there is no rationale why we can neglect
the configuration in which both instantons and
anti-instantons are present. We can only hope that
suitable choice of the model might
take into account such effect implicitly.
A concrete proposal along this line  was
put forward first by Ishibashi, Kawai, Kitazawa and Tsuchiya
\cite{ikkt}.
Their proposal is essentially to regard the above
simple action as the exact action in the following form.
\EQ
Z_{\rm{IKKT}}= \sum_N \Bigl(\prod_{\mu}\int \, d^{N^2}X_{\mu}
\Bigr) d^{16}\psi
\exp S[X, \psi, \alpha, \beta]
\EN
\EQ
S[X, \psi, \alpha, \beta] \equiv  -\alpha N +
\beta \Tr_N{1\over 2}[X_{\mu}, Y_{\nu}]^2
- \cdots
\label{ikkt}
\EN
with $\alpha \propto {1\over g_s\ell_s^4}, \beta \propto {1\over g_s}$
and the summation over $N$ is assumed.
 Indeed, it has been shown that the model
is consistent with the leading long-distance behavior of
D-brane interactions expected from the
low-energy effective theory, the type IIB supergravity in 10 dimensions.

Is there any signature of the space-time uncertainty
principle in this model? Clearly, we cannot extend the
previous scaling arguments to the present model, since
there is no world volume coordinate.
The crucial difference is that now the time
is also treated as a matrix. Formally, the type IIB
matrix action can be regarded as the T-dual correspondent to
the type IIA matrix action. Hence, the
scaling of the time matrix is now reversed from the
time coordinate of the latter.  Thus the scaling
property simply leads to the 10 dimensional
Planck length $g_s^{1/4}\ell_s$ to be
the characteristic scale for D-instantons.
However, D-instantons themselves can never be
used as probes for  the space-time structure and thus
the physical scales can only be determined by the
complicated dynamics of the collective modes
of infinitely many systems of D-instantons.

If we remember the analogy with ordinary quantum mechanics,
it is now tempting to suppose that, instead of the
scaling property, some operator algebra is
the right tool of discussing the space-time
uncertainty principle in the type IIB model, since all
the space-time coordinates are treated as
(infinite) dimensional matrices.
Such an interpretation is proposed in \cite{yo6},
suggesting that the model should be
formulated as a microcanonical
statistical system with a constraint on the
commutation relations of the coordinate matrices
in the form
 \EQ
\langle {1\over 2}([X_{\mu}, X_{\nu}])^2 \rangle \equiv
\lim_{N\rightarrow \infty}
{1\over N}\Tr {1\over 2}([X_{\mu}, X_{\nu}])^2=\alpha'^2 .
\label{matrixqcondition}
\EN
 which leads to the inequality
\EQ
\sqrt{-\langle ([X^0, X^i])^2\rangle } \, \, \ge \alpha',
\EN
in conformity with the space-time uncertainty relation
$\Delta T\Delta X\ge \alpha'$.
The integration measure is determined to be that of
the maximal supersymmetric model, namely the
fermionic determinant,  if we require the
cluster decomposition property of space-time.
 Because of the equivalence
of the canonical and microcanonical ensemble,
we can then
derive the IKKT model using the standard arguments
of statistical mechanics. For more motivations supporting
this interpretation and the relation with
conformal invariance, we refer the reader to ref.  \cite{yo6}.

Now that we have summarized
the space-time uncertainty principle in string theory,
we again turn to the problem of black holes.

\section{Matrix black holes}

Much progress has been made in understanding microscopic origin of
Bekenstein-Hawking entropy in the past few years  \cite{strovafa} (for 
reviews see \cite{maldacena1})
from the standpoint of string theory. Since most
work were based on the physics of D-branes, it has been hard to go
beyond the case of
extremal black holes. Matrix theory provides a unique opportunity
to understand the standard Schwarzschild black holes, as it
purports to provide a nonperturbative formulation for M theory.
As we have seen in the last section, it also forms a natural ground
for space-time uncertainty relation. We will see in the next
section how the two
seemingly different issues are naturally tied up together.

By boosting a hole with
an extremely large longitudinal momentum, one effectively puts
the hole to a background of a large number of D0-branes which
are BPS states. Now the properties of the quantum black hole is
determined by the statistical mechanics of a large number of
D-partons. The geometry is probed by sending in a classical probe
which also consists of a large number of D-partons.

The first observation, due to Banks et al. \cite{bfks}, is that
for a finite
longitudinal cut-off $R$ and a black hole of radius $r_s>R$,
it is necessary to boost the hole in order to fit it into the
asymptotic box size $R$. Asymptotically, one can apply the Lorentz
contraction formula $r_se^{-\alpha}$, where $\alpha$ is the rapidity
parameter, roughly equal to $M/P_-$. The minimal boost is
determined by $r_se^{-\alpha}=R$, or $P_-=r_sM/R$. In matrix
theory, $P_-=N/R$, where $N$ is the number of partons.
The above formula says that $N=r_sM\sim S$, where $S$ is the entropy
of the hole. This condition then says that the minimal number
of partons required to account for entropy $S$ is just $S$,
a physically appealing claim.

Geometrically, one might wonder how the Lorentz contraction could
happen to a horizon, since by definition horizon is a null surface
which is independent of the coordinates used. Indeed it can be shown
that in the boosted frame, the size of horizon remains the same.
What the boost does to the black hole is to change the relation
between the size of the horizon and the asymptotic radius of the
longitudinal direction, if the hole is put on a periodic circle.
It can be shown that for the horizon size to be $r_s$ while the
asymptotic box size to be $R$, the hole must carry a minimal
momentum as determined naively in the last paragraph
\cite{horomarti}.

Another point we want to emphasize here is that when the size of
the hole fits the box size, it looks more like a black string.
Indeed, a black string becomes unstable at the special point
$N\sim S$. Since the horizon area of the black hole of the same
size and same momentum is greater than that of the black string
when one slightly increases the momentum, the black string will
collapse to a black hole.

We will be able to explain the size of the hole and its entropy only
up to a numerical coefficient, thus whenever we write down a formula
that is valid only up to a numerical coefficient. In D dimensional
space-time, the size of the Schwarzschild black hole and its entropy,
written in terms of the mass are given by
\EQ
r_s^{D-3}=G_DM, \quad S={r_s^{D-2}\over G_D}
=G_D^{{1\over D-3}}M^{{D-2\over D-3}}.
\label{basic}
\EN

At the special kinetic point $N\sim S$, we use the second relation
in (\ref{basic}) to solve $M$ in terms of $N$:
$$M=G_D^{-{1\over D-2}}N^{{D-3\over D-2}},$$
thus the light-cone energy
\EQ
E_{LC}=RG_D^{-{2\over D-2}}N^{{D-4\over D-2}},
\label{lce}
\EN
and the size of the hole
\EQ
r_s=(G_DN)^{{1\over D-2}}.
\label{schs}
\EN

As we argued before, the boosted black hole at the transition point
$N\sim S$ can be either regarded as a black string, if the longitudinal
momentum is slightly smaller than the critical value, or a black hole
if the longitudinal momentum is slightly larger. In the former case,
one needs to excite longitudinal objects such as longitudinally
wrapped membrane in matrix theory, thus the momentum modes in the
low energy nonabelian field theory are relevant. Actually the hole
phase is easier to account for. Only the zero modes, in other words,
the motion of D0-branes in the open space are relevant.

When the Born-Oppenheimer approximation is valid, the one-loop,
spin-independent potential between two D0-branes is given in
\cite{dkps}. The
assumption that the Born-Oppenheimer approximation is valid for a
black hole implies that the dominant part of the black hole is a
gas of D0-branes, such that for dynamic purposes one can integrate
out off-diagonal variables. In D dimensional space-time, when M
theory is compactified on a torus $T^{11-D}$, the analogous
potential between two D0-branes can be obtained from that in 11D by
summing over infinitely many images on the covering space of the
torus:
\EQ
L={1\over 2R}(v_1^2+v_2^2)+{c_DG_D\over R^3}{(v_1-v_2)^4
\over r^{D-4}},
\label{totl}
\EN
if the exchange of supergraviton producing the potential does not
cause longitudinal momentum transfer. The above form is certainly
appropriate for D0-branes, and for threshold bound states of D0-branes.
It is a simple matter to use the above interaction and the Heisenberg
uncertainty relation to derive formulas in (\ref{lce}) and (\ref{schs})
\cite{horomarti}\cite{bkfs2}.

However, it is rather bizarre to consider a black hole whose size
is comparable to the size of one spatial dimension of the universe.
To lift the restriction, we need to consider a highly boosted
black hole such that its longitudinal size is much smaller than
its transverse size and has nothing to do with $R$.
In such a situation, we postulate that the black hole is better viewed as a
gas of $S$ clusters whose average longitudinal momentum is $p_-=N/(SR)$
\cite{limarti}.
For the time being $S$ has nothing to do with entropy.
The interaction energy for zero longitudinal momentum exchange
is given by the above formula with $1/R^2$ replaced by $p_-^2$.
Here we need to take one step further, to assume that for processes
in which longitudinal momentum transfer occurs the interaction
takes the more or less the same form, then the total interaction
energy of the gas is
\EQ
V \sim G_D\sum_{\delta p_-=0}
^{N/(SR)}\sum_{a,b}{p_-v^2p_-v^2\over Rr_s^{D-4}}.
\label{suml}
\EN
where the maximal longitudinal momentum exchange is limited by
$p_-=N/(SR)$ too.

The velocity of a cluster is bounded by the Heisenberg uncertainty
relation $p_-v\sim 1/r_s$, thus the interaction energy is bounded
by
\EQ
V\sim G_D{N\over S}S^2{S^2R\over N^2r_s^D}
\sim E_T{G_DS\over r_s^{D-2}}.
\label{poten}
\EN
The virial theorem $V\sim E_T$ implies that $S\sim r_s^{D-2}/G_D$, the
Bekenstein formula. Here we assumed that the number of clusters determines
entropy.

Note that the above is the {\it lower} bound for the
interaction energy, according to Heisenberg relation, thus the
Bekenstein-Hawing formula is a {\it upper} bound. This explains
two interesting points.
\begin{enumerate}
\item If the localization size of a cluster is
$r<r_s$, then we would obtain a formula $S\sim r_s^{D-2}r^2/G_D$
which is smaller than the hole entropy.
\item Even when the localization
size is $r_s$, the Heisenberg relation must be saturated in order
to attain the maximal entropy given by the Bekenstein-Hawking formula.
\end{enumerate}
As argued in \cite{li}, it is also possible to account for the correct
virial theorem using spin-dependent interactions among individual
D0-branes.

It can be shown that the contribution to the potential energy from
other forms of interaction is the same order as (\ref{poten}).
The fact that there are about $S$ clusters suggests that these
clusters obey Boltzmann statistics. This is easy to justify
for large $N$, since each cluster may have some fluctuation
in its longitudinal momentum. It is also possible that some
background whose kinetic energy is negligible is responsible
for the distinguishable clusters.

The correspondence principle of Horowitz and Polchinski, 
as explained already in the previous section,  can be
derived, using the above information from matrix theory. Furthermore.
the dynamic details of string/hole transition can be studied
in the matrix string picture \cite{limartisaha}. For related
works, see \cite{coll}.

We now recall what we have learned.
Under a large boost, the longitudinal size of the black hole
as seen by a distant observer contracts enormously. Denote this size
by $r_L$, then its relation to the Schwarzschild radius is
$r_L=r_se^{-\alpha}=r_sM/P_-$. Again to a distant observer, the
black hole is described as a gas of clusters each carrying
longitudinal momentum $p_-\sim 1/r_L\sim P_-/(r_sM)=P_-/S$.
This gas is best viewed as a version of the quantum stretched
horizon, and the fact that there are $S$ such clusters strongly
suggests a holographic picture, namely the black hole physics
is effectively described by clusters in a $D-1$ dimensional theory.
We shall see in the next section, that this picture also motivates
a generalization of space-time uncertainty relation to M theory,
whose validity goes beyond string theory.

\section{Space-time uncertainty relation and holography}
\subsection{M-theory interpretation of the space-time
uncertainty relation}

The space-time uncertainty relation $\Delta X\Delta T\ge \alpha'$
has a firm ground in string theory, as we have
argued in section 2  where its possible relevance
as the underlying principle for holography is also
suggested. In the context of matrix theory,
the space-time uncertainty relation takes the form
\EQ
\Delta X_T\Delta T\ge  {l_p^3\over R}
\label{mun}
\EN
where $\Delta X_T$ and $\Delta T$ are the characteristic
scales with respect to
spatial transverse distances and time,
respectively.
This relation is about processes involving
individual D-partons, whose longitudinal wave length is given by
$R$. Admittedly, this is a rather weak relation in a process when
more complicated composite objects are involved.

A reinterpretation
of the above relation from the viewpoint of
M theory allows us to write down a more general relation.
Note that a process involving individual D-partons necessarily
smears over the longitudinal direction, thus the uncertainty in
this direction $\Delta X_L=R$ is maximal. Relation (\ref{mun}) is
rewritten as
\EQ
\Delta X_T\Delta X_L\Delta T\ge l_p^3,
\label{munc}
\EN
this relation
\footnote{
This is reminiscent of a triple uncertainty relation
$\Delta X\Delta Y^2 \ge g_s \ell_s^3$
suggested in \cite{liyo}.
} refers only to the fundamental
length scale in M theory,
the Planck length, thus it is a natural candidate for the generalized
uncertainty relation in M theory. In the previous section, we
learned that in a highly boosted object the natural constituent
is a cluster whose longitudinal wave length does not exceed
the longitudinal size of the macroscopic object. We now argue
that relation (\ref{munc}) is the correct relation for a process
involving such a cluster. It is trivially true for threshold
bound state, since it is just a boosted parton and according to
Lorentz invariance $\Delta X_L$ contracts, while $\Delta T$
is dilated by a same factor. An object carrying the same
amount of longitudinal momentum can be regarded as an excited state
of the threshold bound state, therefore intuitively as a probe
it can not probe a transverse distance shorter than a threshold
bound state can do. Thus, relation (\ref{munc}) must also hold
for such a probe. Note also that this relation is Lorentz invariant.

A good example of a cluster is the one discussed in  \cite{horomarti}.
That is a spinning membrane. If we set its transverse size
$X_T=1$, then the time scale is $T=N/R$. Now since the spinning
membrane carries longitudinal momentum $P_-=N/R$ and so its
longitudinal extension is $X_L=1/p_-=R/N$. The cubic uncertainty
relation is satisfied.

As explained in the last part of section 2,
one of us argued in \cite{yo6} that the string uncertainty relation serves
as a guiding principle in constructing the IIB matrix model.
We suspect that if the M theory uncertainty relation (\ref{munc})
has any deep meaning, it may also underlie the ultimate
nonperturbative and covariant
formulation of M theory whose validity goes
beyond Matrix theory which is only formulated in the light-
cone frame (or IMF).

In the AdS/CFT correspondence conjectured by Maldacena
\cite{maldacena2}, the
holographic bound is satisfied provided one notices the UV/IR
correspondence between a bulk process and a process in the
boundary theory. To see how our uncertainty relation (\ref{munc})
fits into this context, we examine the $AdS_7\times S^4$ case.
The reason for picking out this one is obvious: There is
no underlying string theory, thus relation (\ref{munc}) is
most suitable. The metric of the $AdS_7$ is given by
\EQ
ds^2={r\over R}(-dt^2+\sum_{i=1}^5dx_i^2)+{R^2\over r^2}dr^2,
\label{metr}
\EN
where the coordinates $(t, x_i)$ are the natural ones on fivebranes,
we have set $l_p=1$.

The simplest way to probe this background is through the use
of open membranes. An open membrane appears as a string inside
fivebranes. Our uncertainty relation says that if the
uncertainty in the brane theory is given by $\Delta T$ and
$\Delta X_L$,
then necessarily the membrane state
will have an extension in the
transverse direction (the radial direction) at least
$\Delta X_T\sim 1/(\Delta T\Delta X_L)$. This translates
into a statement in terms of coordinates in (\ref{metr})
that $\Delta r\sim 1/(\Delta t \Delta x)$, thanks to the
peculiar form of the metric. It is easy to see that
this is a  generalization of the UV/IR correspondence
observed by Susskind and Witten, as discussed in the
next subsection.

A further support to the above relation is the scaling argument
similar to the one for the original
uncertainty relation for strings and D-branes.
$r$ naturally corresponds to scalars
on fivebranes. There are 5 scalars on each fivebrane representing
transverse fluctuations, in a tensor super-multiplet. Now the scalar
in the 6D conformally
invariant theory has dimension 2, namely if $x_i\rightarrow
\lambda x_i$, then $\phi\rightarrow \lambda^{-2}\phi$. This
is precisely our stated uncertainty relation.

The generalized space-time
uncertainty relation says nothing about the nature of the
theory living on the fivebranes, except for the above 
scaling property. As observed in \cite{susswitten}, so
long if one has a consistent short distance theory on the branes,
the theory is able to account the entropy in the whole bulk
space. This is certainly a good qualitative argument for
Maldacena's conjecture. To obtain the right Bekenstein-Hawking
bound in this particular example, we need to use another
set of coordinates, similar to those in \cite{susswitten}.
As pointed out
by Banks, the number of degrees must be proportional to $N^3$ in
this case, in other words, the energy density must be proportional
to $N^3/\delta^6$, $\delta$ is the UV cut-off. The close relation
between holography and spacetime uncertainty relation certainly
suggests that the latter should have a deep meaning. 
Related views on the interpretation of the space-time uncertainty principle and its relation to holography are 
independently pointed out by Minic \cite{minic}.  

\subsection{The space-time uncertainty relation and the
UV/IR correspondence of
Susskind-Witten}
Finally, we want to emphasize that the UV/IR correspondence
underlying Susskind-Witten's  derivation \cite
{susswitten} of the holographic
bound for the entropy of the d3-brane is regarded as a
consequence of the string space-time uncertainty relation. 
Following \cite{susswitten}, let us 
assume the infrared cutoff of the
bulk theory to be $A^{1/3} \sim R/\delta$
with respect to the volume $A$ of the boundary.
Here we use the corrected metric ($R=\ell_s(g_sN)^{1/4}$)
for the $AdS_5$ space
\EQ
ds^2 = R^2\Bigl[{4dx^i dx^i \over (1-r^2)^2} -dt^2 {(1+r^2)^2 \over
(1-r^2)^2}\Bigr].
\label{sw-metric}
\EN
This cutoff is equivalent to introducing the
uncertainty $\Delta r \sim \delta$ in the radial (
{\it i.e.}, transverse) direction
orthogonal to the boundary $r=1$.  In terms of the
proper distance, $\Delta X \sim R\delta/\delta =R$.
Then by the string space-time
uncertainty relation,
we have the corresponding cutoff
$\Delta T \sim \ell_s^2/R$ for the proper time,
 or, in terms of the coordinate time, 
$\Delta t \sim \ell_s^2\delta/R^2$.  If we
use the dimensionless unit of ref. \cite{susswitten}
for the surface theory,
we have $\Delta t \sim \delta$.  This explains the UV cutoff
for the boundary theory adopted in \cite{susswitten}, where
the coordinates of the boundary theory is assumed to be
$(t, x^i)$ with $x^ix^i \sim 1$, which leads to the
number of degrees of freedom $N_{dof}=N^2/\delta^2
=A/G_5$. 
It is easy to extend the arguments to M2 and M5 branes
by using appropriate coordinates similar to (\ref{sw-metric})
and determine the $N$ dependencies required for
agreement with the Beckenstein-Hawking formula.
For M5 with $R\sim \ell_P N^{1/3}$ and
M2 with $R\sim \ell_PN^{1/6}$,
we need $N_{dof}\sim N^3/\delta^5$ and
$N_{dof} \sim N^{3/2}/\delta^2$, respectively.

It should be kept in mind that,  because of the different
choice of the coordinates, the uncertainty relation
for the coordinate time and coordinate
spatial distance
now takes the form $\Delta t \Delta r \sim \ell_s^2\delta^2/R^2$
which depends on the cutoff $\delta$.
 In the metric of the type (\ref{metr})
(or (\ref{decouplinglimit}) for D0 case), on the other hand,
the (generalized) uncertainty
relation is valid in the same forms  for the proper
distances and the coordinate distances.
For other choices of coordinates, we must be careful
in  applying the uncertainty relation. 
In view of the original derivation of the 
space-time uncertainty relation, as explained in subsection 2.1,
we have to use the proper distances for physical
space-time lengths, $\Delta X$ and $\Delta T$.

\section{Conclusion}
We have reviewed the present status of string theory
from the viewpoint of the short-distance space-time
structure and black-hole entropy. In particular,
we have mainly emphasized the space-time
uncertainty relation as a possible guiding principle
for understanding some universal short-distance
structure in perturbative and non-perturbative
string and M theories. We have also suggested
the relevance of the space-time uncertainty relation
for holography.

There are of course other viewpoints in aiming towards
the same goal. Besides some related approaches
already mentioned,  we have not emphasized, for example,
 the viewpoint of noncommutative geometry as
a possible mathematical framework for formulating the
noncommutative structure of space-time geometry
at short distances. For an example of
such approaches, we refer the reader
to \cite{douglas}\cite{douglas2}
and to different attempts such as 
\cite{kato}\cite{oda}.
We hope that our physical and intuitive
viewpoint would play  a  useful role which is complementary to
 more abstract and formal investigations.
In particular, we expect that one of the
most intriguing questions is how to unify
various conformal symmetries, 
which are now turned out to be the formal bases for
 the space-time uncertainty principle,   and how to incorporate
them into a fundamental mathematical formalism.

\vspace{1cm}
\noindent
{\it Acknowledgments}

In preparing the present article, we have benefited
from conversations and e-mail exchanges
with A. Jevicki, M. Kato,
Y. Kazama, H. Kawai,  E. Martinec, D. Minic.
The work of M. L. is supported by DOE grant DE-FG02-90ER-40560 and NSF grant
PHY 91-23780.
The work of T. Y. is supported in part by
Grant-in-Aid for Scientific  Research (No. 09640337)
and Grant-in-Aid for International Scientific Research
(Joint Research, No. 10044061) from the Ministry of  Education, Science and 
Culture.

\end{document}